\documentclass[12pt]{article}
\usepackage{amsmath}
\usepackage{amsfonts}
\usepackage{amsthm}
\usepackage{amssymb}
\usepackage{graphicx}
\usepackage{enumerate}
\usepackage{natbib}
\usepackage{url} 
\usepackage{graphicx}
\usepackage{caption}
\usepackage{subcaption}
\usepackage{comment}
\usepackage{booktabs}
\usepackage[table,xcdraw]{xcolor}
\usepackage{tikz}
\usepackage{hyperref}
\usepackage{algorithm}
\usepackage{algpseudocode}
\usepackage{float}
\usepackage{import}
\usepackage{caption}     
\usepackage{xr}
\newcommand{\blind}{1}

\begin{document}

\def\spacingset#1{\renewcommand{\baselinestretch}%
{#1}\small\normalsize} \spacingset{1}


\if1\blind
{
\title{\bf Interval Estimation for Binomial Proportions Under Differential Privacy}

\author{
  Hsuan-Chen (Justin) Kao\thanks{justinkao.44@duke.edu, 214 Old Chemistry, Box 90251, Durham, NC 27708-0251, USA} \footnotemark[3] \quad
  Jerome P. Reiter\thanks{jreiter@duke.edu, 214 Old Chemistry, Box 90251, Durham, NC 27708-0251, USA} \\
  Department of Statistical Science, Duke University
  }

\renewcommand{\thefootnote}{\fnsymbol{footnote}}
\footnotetext[3]{The authors gratefully acknowledge support from NSF SES 2217456.}

  \maketitle
} \fi

\if0\blind
{
  \bigskip
  \bigskip
  \bigskip
  \begin{center}
    {\bf Interval Estimation for Binomial Proportions Under Differential Privacy}
\end{center}
  \medskip
} \fi

\bigskip
\begin{abstract}
When releasing binary proportions computed using sensitive data, several government agencies and other data stewards protect confidentiality of the underlying values by ensuring the released statistics satisfy differential privacy.  Typically, this is done by adding carefully chosen noise to the sample proportion computed using the confidential data. In this article, we describe and compare methods for turning this differentially private proportion into an interval estimate for an underlying population probability. Specifically, we consider differentially private versions of the Wald and Wilson intervals, Bayesian credible intervals based on denoising the differentially private proportion, and an exact interval motivated by the Clopper-Pearson confidence interval. We examine the repeated sampling performances of the intervals using simulation studies under both the Laplace mechanism and discrete Gaussian mechanism across a range of privacy guarantees. We find that while several methods can offer reasonable performances, the Bayesian credible intervals are the most attractive. 
\end{abstract}

\noindent%
\vfill

\newpage

\section{Introduction}\label{chap:Introduction}

Many government agencies, researchers, and other organizations---henceforth all called agencies---consider sharing data with the public  a key part of their missions  \citep{reiter2019differential}. These agencies also typically have to protect the confidentiality of data subjects' identities and sensitive attributes. 
However, protecting confidentiality  is challenging due to the proliferation of readily available digital data and analytical tools that could help adversaries make disclosures from the information released by the agency.  In fact, even releasing summary statistics like proportions or counts has been shown to introduce disclosure risks, as given enough of these statistics adversaries may be able to reconstruct the underlying confidential data \citep{dinur:nissim, dwork:exposed, census:recon}.

As a result of these risks, several agencies now release statistics that satisfy differential privacy (DP) \citep{dwork2006calibrating, dwork2006differential}, including the Bureau of the Census and Internal Revenue Service in the U.\ S.\, as well as companies like Apple \citep{apple2017learning}, Uber \citep{208167}, Microsoft, Meta, and Google. DP is a mathematical criterion that ensures the released statistics are not overly sensitive to the inclusion or exclusion of any particular individual in the underlying confidential data. A typical way to implement DP is to add carefully calibrated noise to the statistic computed using the confidential data.  The noisy statistic is released to the public, along with a description of the distributions used to generate the noise.

When evaluating the efficacy of DP algorithms, generally researchers compare the released statistic to the corresponding statistic computed with the confidential data, making statements about how far apart the two are likely to be according to the noise distribution.  Often they do not consider whether the DP algorithm can generate inferences about the underlying data generating process or population parameter. There are notable exceptions, for example, methods for significance testing \citep[e.g., ][]{vu:slav, DPchisq, awan2020differentially, Wang2015RevisitingDP} 
and interval estimation  \citep[e.g., ][]{honaker, karwa, awan2020differentially, Covington_2024, lin:bun, li:reiter}.


In this article, we describe and compare several interval estimates for binomial proportions under DP.  Specifically, we consider using 
the differentially private proportions in the expressions for the usual Wald and Wilson confidence intervals.   
We consider Bayesian credible intervals constructed  from the posterior distribution of the true proportion given the noisy proportion. We examine Bayesian intervals using both a uniform prior distribution and a Jeffreys prior distribution. 
Finally, we construct an exact interval following the strategy used in the formation of the usual Clopper-Pearson interval absent privacy considerations.  
 We present intervals for two types of DP definitions, namely pure-DP and  R\'enyi DP \citep{DBLP:journals/corr/Mironov17}, using a Laplace mechanism for the former and a 
discrete Gaussian  mechanism for the latter. Using simulation studies, we examine the repeated sampling properties of these intervals under different privacy guarantees.  The simulation results suggest that several of the methods offer reasonable performance, and that arguably the Bayesian credible intervals are the most attractive. 

The rest of this article is organized as follows. Section \ref{chap:Background} briefly reviews the usual Wald and Wilson intervals for binomial proportions absent privacy concerns. It also summarizes the two variants of differential privacy that we utilize, namely pure DP and R\'enyi DP. Section \ref{chap:failure} highlights problems that can arise if one constructs intervals by inserting  
the DP estimate of the sample proportion in expressions for the Wald and Wilson intervals. Section \ref{alternative} describes the Bayesian credible intervals based on denoising the differentially private proportion 
and the exact interval motivated by the Clopper-Pearson confidence interval. Using Bayesian inference for interval estimation with proportions has been suggested previously in the DP literature \citep[e.g.,][]{li:reiter}, but we believe the 
exact intervals have not been proposed previously.  
Section \ref{sec:sims} describes the results of the repeated-sampling simulation studies. Here, 
our purpose is to evaluate the performance of the interval estimates within each DP paradigm; we do not compare performances of the intervals  across pure DP and R\'enyi DP. Finally, Section \ref{sec:conc} concludes with some suggestions for topics for further investigation.

\section{Background}
\label{chap:Background}

We first review the Wald and Wilson intervals, followed by the review of DP including
the Laplace mechanism and discrete Gaussian mechanism.

\subsection{Wald and Wilson Intervals}\label{sec:waldwilson}

Suppose we have a sample of independent and identically distributed data, $(x_1, \dots, x_n)$, where  $x_i \in \{0, 1\}$  for $i=1, \dots, n$.  Let $X=\sum_{i=0}^nx_i=k$ be the random variable representing the number of successes out of the $n$ random trials that could have been observed, so that $X \sim \textsf{Binomial}(n, p)$, where $p$ is the probability of success. Let $\hat{p} = \sum_{i=1}^n x_i/n$ be the sample proportion.  


Two well-known methods to construct interval estimates for the unknown $p$ include the Wald and Wilson intervals. The Wald interval is
\begin{equation}
\hat{p} \pm z_{\alpha / 2} \sqrt{\hat{p}(1-\hat{p})/n},\label{eq:wald}
\end{equation}
where $z_{\alpha / 2}$ is the $(1 - \alpha/2)$ quantile from the standard normal distribution, where commonly $\alpha = 0.05$.
It is mathematically possible for the upper bound of \eqref{eq:wald} to exceed one or the lower bound of \eqref{eq:wald} to fall below zero. Further, \eqref{eq:wald} returns a single value when $\hat{p}=0$ or $\hat{p}=1$.  Lastly, the Wald interval relies on the central limit theorem holding, which may not be the case for some $n$ \citep{Wallis01082013}.

The  Wilson interval is
\begin{equation}
\left( \frac{1}{1 + z_{\alpha/2}^2/n} \right)\left(\hat{p} + \frac{z_{\alpha/2}^2}{2n} \pm z_{\alpha/2} \sqrt{\frac{\hat{p}(1-\hat{p})}{n} + \frac{z_{\alpha/2}^2}{4n^2}}\right).\label{eq:wilson}
\end{equation}
  In contrast to the Wald interval, the interval from \eqref{eq:wilson} is guaranteed to lie inside [0,1]. It produces an interval when $\hat{p}=0$ or $\hat{p}=1$.  
It also can have slightly better confidence interval coverage properties, as evident in simulation studies reported in the literature \citep[e.g., ][]{Wallis01082013, 10.1214/ss/1009213286, newcombe1998two, lottreiter2020wilson}.  For these reasons, some researchers recommend the Wilson interval over the Wald interval, although the latter remains widely used in practice.  



\subsection{Differential Privacy}\label{sec:DPreview}

Differential privacy has become a gold-standard definition of what it means for data products to be confidential. DP depends on the concept of neighboring databases. In our context, we consider the common definition of two neighboring databases, say $D$ and $D'$, as differing on only one observation.  For example, we could have $D'= D \cup x$, where $x$ is some other value in the domain of the variable of interest. Alternatively, we could have $D'= \{D - x_i\} \cup x$ for some $x_i \in D$ and some $x$; that is, $D'$ is constructed by replacing one of the values in $D$ with some value $x$ in the domain of the variable of interest.  The latter definition of neighboring databases presumes both $D$ and $D'$ have the same sample size, which implies that the sample size $n$ can be considered known.  We presume this definition of neighboring databases, as we use  $n$ to construct interval estimates for $p$.

\subsubsection{Pure DP and the Laplace mechanism}\label{sec:pureDP}

Let $\mathcal{M}$ be some algorithm that takes any dataset $D$ as an input and produces an output in some set $S$ with probability $P(\mathcal{M}(D) \in S)$. We say that $\mathcal{M}$ satisfies $\epsilon$-DP, also called pure DP of just DP for brevity, if, for all neighboring databases $D$ and $D'$ and any output set $S$, we have  
\begin{equation}
P(\mathcal{M}(D) \in S) \leq e^{\epsilon} P(\mathcal{M}(D^{\prime}) \in S).\label{eq:DP} 
\end{equation}
The parameter $\epsilon$ controls the level of privacy offered by $\mathcal{M}.$  When $\epsilon$ is small, one cannot easily tell whether any particular output $S$ was generated by $D$ or $D'$.  Hence, adversaries cannot easily learn the presence or absence of any single record in the confidential data, $(x_1, \dots, x_n).$  When $\epsilon$ is large, the privacy guarantee is less stringent.  The DP literature recommends values of $\epsilon$ around one or less, although in practice larger values are used, as they typically result in $\mathcal{M}$  with smaller noise variance and hence greater accuracy \citep{kazan:reiter}. 

DP has some appealing features \citep{dwork:roth}.  First, it satisfies composition.  If $\mathcal{M}_1$ satisfies $\epsilon_1$-DP and $\mathcal{M}_2$ satisfies $\epsilon_2$-DP, then applying both $\mathcal{M}_1(D)$ and $\mathcal{M}_2(D)$  satisfies $(\epsilon_1+\epsilon_2)$-DP.  Second, it satisfies parallel composition.  If we have two databases $D_1$ and $D_2$ such that $D_1 \cap D_2 = \emptyset$, then applying $\mathcal{M}_1(D_1)$ and $\mathcal{M}_2(D_2)$ satisfies $\max(\epsilon_1, \epsilon_2)$-DP.  Third, it satisfies post-processing.  If we apply any nontrivial function $g$ to the output produced by an $\epsilon$-DP algorithm $\mathcal{M}(D)$, then $g(\mathcal{M}(D))$ also satisfies  $\epsilon$-DP.  We leverage the post-processing property in particular when constructing the interval estimates for $p$.



Many algorithms for implementing DP, including those we use here, rely on a quantity called the global sensitivity. Let $h$ be some function that we wish to apply to $D$, resulting in an output $h(D)$. For example, $h$ could compute the sample proportion of the observed data.  We define the global sensitivity $\Delta_h$ to be the maximum amount that $h$ can change over all possible neighboring databases $(D, D').$ For example, when $h$ is the function that computes the sample proportion of $n$ binary values, $\Delta_h = 1/n$. 

\citet{dwork2006calibrating} show that one can satisfy DP using the Laplace mechanism. This algorithm adds random noise to $h(D)$, where the noise is sampled from a Laplace distribution centered at zero with variance scaled according to $\epsilon$ and $\Delta_h$. For the sample proportion, the Laplace mechanism results in 
    $\hat{p}^* = \hat{p} + \eta$,
where $\eta \sim \textsf{Lap}(0, \Delta_h / \epsilon)$.  The agency releases  $\hat p^*$, possibly truncated to zero or one to enhance face validity of the released proportion.



\subsubsection{R\'enyi DP and the discrete Gaussian mechanism} \label{sec:renyi}

A popular variant of differential privacy is R\'enyi differential privacy  \citep{DBLP:journals/corr/Mironov17}. Let $\mathcal{M}$ be a randomized algorithm that produces an output distribution
$P_{\mathcal{M}(D)}$ when applied to a dataset $D$.  For neighboring datasets $D$ and $D'$, the R\'enyi divergence of order $\lambda > 1$ is defined as
\begin{equation}
    D_{\lambda}\!\big(P_{\mathcal{M}(D)} \,\|\, P_{\mathcal{M}(D')}\big)
= \frac{1}{\lambda - 1}
\log \text{E}_{y \sim P_{\mathcal{M}(D')}}\!\left[\!
\left(\frac{P_{\mathcal{M}(D)}(y)}{P_{\mathcal{M}(D')}(y)}\right)^{\lambda}
\!\right],
\end{equation}
where $P_{\mathcal{M}(D)}(y)$ and $P_{\mathcal{M}(D')}(y)$ are the densities of the two output distributions at any value $y$. We say that $\mathcal{M}$ satisfies $\epsilon$-R\'enyi differential privacy with order $\lambda$ if, for all neighboring datasets $D$ and $D'$, 
\begin{equation}
    D_{\lambda}\!\big(P_{\mathcal{M}(D)} \,\|\, P_{\mathcal{M}(D')}\big) \le \epsilon.
\end{equation}

One way to satisfy R\'enyi DP is to add noise to $f(D)$ using a Gaussian distribution  \citep{DBLP:journals/corr/Mironov17}.  We note that adding noise from a Gaussian distribution also can satisfy what is known as $(\epsilon, \delta)$-DP; we  consider only R\'enyi DP here. The Gaussian mechanism can offer advantages for accuracy over the Laplace, as the Gaussian distribution has a lower chance of generating large values of noise (compared to the Laplace distribution) due to its tail behavior.  We also can 
add noise from a discretized version of the Gaussian distribution that has support over the integers \citep{Canonne_2022}. The probability mass function of this distribution is
\begin{equation}
    P(G=g)=\frac{\exp \left(-\frac{g^2}{2 \sigma^2}\right)}{\sum_{m=-\infty}^{\infty} \exp \left(-\frac{m^2}{2 \sigma^2}\right)}, \quad g \in  \{\dots, -2, -1, 0, 1, 2, \dots\}.\label{eq:DG}
\end{equation}
When releasing a noisy version of the sample proportion, the discrete Gaussian mechanism is given by 
\begin{equation}
\hat p^* = \frac{\sum_{i=1}^n x_i + g}{n}, \quad g \sim \textsf{DiscreteGaussian}\left(0, \sigma^2\right),\label{eq:DGM}
\end{equation} 
Note that we add discrete Gaussian noise to the count. 

Using results from \citet{Canonne_2022} and \citet{DBLP:journals/corr/Mironov17}, together with the fact that the $\ell_1$ sensitivity of the summation function $\sum_{i=1}^n x_i$ is $\Delta_h = 1$, it follows that the mechanism in \eqref{eq:DGM} satisfies $(\lambda, \lambda / (2\sigma^2))$-Rényi differential privacy.

\section{Naive Wald and Wilson Intervals Under DP}
\label{chap:failure}

Under the Laplace mechanism, $E(\hat p^*) = E(\hat{p})=p$, and 
$\text{Var}(\hat{p}^*) = p(1 - p)/n + 2/(n^2 \epsilon^2).$ Thus, one possible interval substitutes $\hat p^*$ for $\hat{p}$ and $\text{Var}(\hat{p}^*)$ for $\text{Var}(\hat{p})$ in the expression for the Wald interval in \eqref{eq:wald}.   This results in what we call the 
Wald interval using DP point estimates, or Wald-DP for short,
given by   
\begin{equation}
\hat{p}^* \pm z_{\alpha / 2} \cdot \sqrt{\frac{\hat{p}^* (1 - \hat{p}^*)}{n} + \frac{2}{n^2 \epsilon^2}}.\label{pluginWald}
\end{equation}
When computing \eqref{pluginWald}, if $\hat p^* <0$ we can set it to zero and if $\hat p^* >1$ we can set it to one  These are post-processing operations and hence do not incur any extra privacy loss.  

Even with this clipping, it is evident from the margin of error in \eqref{pluginWald} that this interval can exacerbate the out-of-bounds problem noted in Section \ref{sec:waldwilson}, especially when $\epsilon$ and $n$ are small. One solution is to clip the Wald-DP interval itself at zero and one, although there is no theory that underpins the repeated-sampling validity of this practice.


Alternatively, we can naively follow the computations used to construct the Wilson interval in \eqref{eq:wilson}.  To do so, we solve for $p$ in
\begin{equation} \label{plug_in_wilson}
    \left( n + z_{\alpha / 2}^2 \right) p^2 - \left( 2 n \hat{p}^* + z_{\alpha / 2}^2 \right) p + \left( n \hat{p}^{* 2} - \frac{2 z_{\alpha / 2}^2}{n \epsilon^2} \right) \leq 0.
\end{equation}
As needed, we clip $\hat{p}^*$ to zero and one when computing \eqref{plug_in_wilson}.
This ad hoc interval, which we call the 
Wilson interval using DP point estimates or Wilson-DP for short, 
presumes that $(\hat p^* - p)^2/ \text{Var}(\hat p^*)$ follows a chi-squared distribution on one degree of freedom.  However, this distributional assumption is incorrect. Nonetheless, we evaluate the resulting interval in Section \ref{sec:sims}. Notably, the Wilson-DP interval can 
fall outside $[0, 1]$. We show this in detail in 
the supplementary material.

\section{Principled Intervals Under \texorpdfstring{$\epsilon$} \ \ -DP}
\label{alternative}
The Wald-DP and Wilson-DP intervals are ad hoc in that they do not fully account for the noise mechanism in a principled manner.  In this section, we present three intervals that do so in different ways.  We mainly present the intervals under $\epsilon$-DP and the Laplace mechanism, although they can be extended to other variants of DP and other algorithms that add noise to the sample proportion.  As an example, we present the Bayesian credible interval for R\'enyi DP and the discrete Gaussian mechanism in Section \ref{alt_renyi}. 

The analyst does not know $p$, of course, nor does the analyst know the confidential value  $\hat p$.  They only have access to $\hat p^*$, as well as a description of the DP algorithm used to create it.  We presume that $n$ is known, e.g., it is provided by the agency.  We also presume that the agency provides the value of $\hat p^*$ without any post-processing, regardless of whether or not it is inside $[0,1].$  Releasing differentially private values without any clipping or other post-processing enables unbiased estimation of $\hat{p}$ and hence $p$. This release strategy is used, for example, by the U.\ S.\ Bureau of the Census, which releases noisy counts without post-processing as part of the 2020 decennial census data products.  We note that the agency could, for convenience, additionally provide a clipped version of $\hat p^*$ with no additional privacy loss.  In Section \ref{sec:conc}, we discuss how to modify the interval estimates when the agency releases only a clipped version of $\hat{p}^*$.  

\subsection{Bayesian Credible Intervals Under \texorpdfstring{${\epsilon}$} \ \ -DP} \label{alt_Bayes_flat}


Following a Bayesian paradigm, let $Q$ be the analyst's random variable for the unknown $p$, and let $q \in [0,1]$ be a realization of $Q$. Let $\hat{Q}$ be the analyst's random variable for the unknown $\hat p$, and let $\hat{q} \in \{0/n, 1/n, \dots, (n-1)/n, n/n\}$ be a realization of $\hat{Q}$.  We seek to estimate the posterior distribution of the unknown $p$ based on the released noisy proportion $\hat{p}^*$. That is,  we estimate the posterior density  
\begin{equation}\label{posterior_of_p}
    f(q \mid \hat{p}^*) \propto f(\hat{p}^* \mid q) f(q). 
\end{equation}
We compute $f(\hat{p}^* \mid q)$ by summing over values of the unobserved sample proportion, 
 using the density
\begin{equation} \label{integral}
    f(\hat{p}^* \mid q) = \sum_{k=0}^n f(\hat{p}^* \mid \hat{q} = k/n) f(\hat{q} \mid q).
\end{equation}
In \eqref{integral}, we use the binomial distribution  
$n\hat{q} \sim \textsf{Binomial}(n, q)$ for $f(\hat{q}|q)$  and the Laplace distribution, 
\begin{equation}
   f(\hat{p}^* \mid \hat{q}) = \frac{n\epsilon}{2} \exp\left(-|\hat{p}^* - \hat{q}|n\epsilon\right). \label{eq:phat}
\end{equation}
Once we have many draws of $q$ from \eqref{posterior_of_p}, we take as the DP interval estimate for $p$ the 2.5 percentile and 97.5 percentiles of these sampled draws.

\subsubsection{Uniform prior} \label{alt_Bayes_uniform}
We first consider the interval based on the uniform prior distribution, $f(q) =  \textsf{Uniform}(0, 1)$. This prior distribution reflects the prior belief that any value of $p$ between 0 and 1 is equally likely. 
The posterior distribution can be sampled in several ways. One approach is to 
use a Gibbs sampling strategy.
The Gibbs sampler alternates between sampling values of $q$ and of $\hat{q}$  from their full conditional distributions. For $q$, 
given the updated value of $\hat{q}$, say $\hat q=k/n$, we have 
$f(q \mid \hat{q}) \propto q^k(1-q)^{n-k}$. Hence, the full conditional distribution of $q$ follows the Beta distribution, 
\begin{equation}\label{eq:fullcondq}
   q \mid \hat{q} \sim \textsf{Beta}(n\hat{q} + 1, n(1 - \hat{q}) + 1). 
\end{equation}
For $\hat{q}$, we apply a grid-based method.  Given the draw of $q$, we evaluate 
\begin{equation}
    f(\hat{q} \mid q, \hat{p}^*) \propto \exp\left(-|\hat{p}^* - \hat{q}|n \epsilon\right) q^{n\hat{q}} (1 - q)^{n(1 - \hat{q})}.\label{eq:likelihood}
\end{equation}
over the grid $\hat{q} \in \{0, 1/n, 2/n, \dots, 1\}$.
We normalize these values to form a probability distribution, from which we sample to get the updated draw of $\hat q$. 

Alternatively, we can approximate \eqref{posterior_of_p} for a finite but large set $Q_0$ of possible values of $q$ that span $[0,1]$, e.g., $Q_0 = \{0, .001, .002, \dots, .999, 1\}.$  Specifically, for each element  $q \in Q_0$, we compute \eqref{integral}.
We then normalize these values to form an approximate posterior distribution for $p$ defined over the values of $q \in Q_0$.  
An advantage of this grid sampler is that it avoids Markov chain Monte Carlo computations, although the support of the approximation is only over the finite set of values of $p$ in $Q_0$.

Finally, one can use numerical integration or  Monte Carlo methods like  rejection sampling, as it is possible to compute the normalizing constant in \eqref{posterior_of_p} under the uniform prior distribution. For brevity, we do not describe other approaches here.

\subsubsection{Jeffreys prior} \label{alt_Bayes_jeffreys}

In the nonprivate setting, the Jeffreys prior is often used for interval estimation,   
as it has the desirable property of reparameterization invariance \citep{10.1214/ss/1009213286,  zanellabéguelin2022bayesianestimationdifferentialprivacy}. 
It corresponds to the Beta prior distribution, 
    $q \sim \textsf{Beta}(1/2, 1/2)$.
Combining the binomial likelihood and the prior distribution, we have 
\begin{equation}
    \begin{aligned}
    f(q \mid \hat{q}) & \propto q^k(1-q)^{n-k} q^{-1 / 2}(1-q)^{-1 / 2} 
     = q^{k-1 / 2}(1-q)^{n-k-1 / 2}, 
\end{aligned}
\end{equation}
which is the kernel of the Beta distribution
\begin{equation}
\begin{aligned}
    q \mid \hat{q} &\sim \textsf{Beta}(k+1 / 2, n-k+1 / 2). \label{eq:postp}
\end{aligned}
\end{equation}
With the Jeffreys prior, we can modify the Gibbs sampler in  Section \ref{alt_Bayes_uniform} to instead sample $q$ given $\hat{q}$ from \eqref{eq:postp}. Likewise, we can define a grid sampler by using \eqref{eq:postp} in lieu of the binomial likelihood in \eqref{integral}. We note that the Jeffreys prior results in a closed form for the normalizing constant, enabling computation of the posterior distribution without the need for Markov chain Monte Carlo sampling.



\subsection{Exact Interval Under \texorpdfstring{$\epsilon$} \ \ - DP}
\label{alt_exact}

We next present an interval motivated by the strategy used to derive the Clopper-Pearson interval. The basic idea is to find the values of $p$ that could give rise to the observed $\hat p^*$ with at least $1 - \alpha$ probability, as we now describe.


We begin by setting a range of candidate values for $p$, covering the interval from 0 to 1 in many small increments.  For each candidate $p$, say $p_j$ where $j=1, \dots, J$ where $J= 1000$, we simulate a sample proportion $\hat{p}_{js} = X_{js}/n$. where $X_{js} \sim \operatorname{Binomial}(n, p_j)$.
We then mimic the Laplace mechanism and add noise $\eta_{js}$ to $\hat{p}_{js}$ drawn from a Laplace distribution with mean zero and scale parameter $1/(n \epsilon)$. This process results in a draw of the noisy sample proportion, $\hat{p}_{js}^*=\hat{p}_{js}+\eta_{js}$.  We repeat the process of generating $\hat p^*_{js}$ many times, say $s=1, \dots, S$ times where $S=5000$, for each $p_j$. The result is a simulated sampling distribution for the noisy proportion for each  $p_j$. 

For each $p_j$, we find the percentages of the $S$ simulated noisy proportions that are greater than or equal to 
$\hat{p}^*$ and that are less than or equal to $\hat{p}^*$.   
We then utilize Brent’s root-finding algorithm \citep{brent1973algorithms}, which we implement via the \textsf{`unitroot()`} function in \textsf{R}, to find bounds $p_L$ and $p_U$. The lower bound $p_L$ is the smallest candidate $p_j$ for which the upper tail probability is less than or equal to $\alpha / 2$, i.e., 
\begin{equation}
p_L=\min_{p_j} \left\{p_j: S^{-1}\sum_{s=1}^S I(\hat p_{js}^* \geq \hat{p}^*) \leq \alpha / 2\right\}.
\end{equation}
The upper bound $p_U$ is the largest candidate $p_j$ for which the lower tail probability is less than or equal to $\alpha / 2$. 
\begin{equation}
p_U=\max_{p_j} \left\{p_j: S^{-1}\sum_{s=1}^S I(\hat p_{js}^* \leq \hat{p}^*) \leq \alpha / 2\right\}.
\end{equation}


\subsection{Bayesian Interval Under R\'enyi DP}
\label{alt_renyi}

We now consider the Bayesian interval under R\'enyi DP and the discrete Gaussian mechanism from Section \ref{sec:renyi}. 
As before let $\hat q \in \{0, 1/n, \dots, (n-1)/n, 1\}$ be a possible value of $\hat Q$, i.e., the random variable representing the unobserved sample proportion.  We seek the posterior distribution $f\left(q \mid \hat{p}^*\right) \propto f\left(\hat{p}^* \mid q\right) f(q) = \sum_{\hat{q}} f\left(\hat{p}^* \mid \hat{q}\right) f(\hat{q} \mid q) f(q)$. We presume a uniform prior distribution for $f(q)$.


Using the discrete Gaussian mechanism from \eqref{eq:DG} and \eqref{eq:DGM}, given a value of $\hat q$, we have
\begin{equation}
    f\left(\hat{p}^* \mid \hat{q}\right) \propto \exp \left(-\frac{\left(n\hat{p}^*-n\hat{q}\right)^2}{2\sigma^2}\right).
    \label{DGp}
\end{equation}
Hence, combining the likelihood in \eqref{DGp} and a uniform prior, the kernel of $f(q|\hat{p}^*)$ is
\begin{equation}
\begin{aligned}
    f\left(q \mid \hat{p}^*\right) &\propto \sum_{k=0}^n \exp \left(-\frac{\left(n\hat{p}^*- k\right)^2}{2 \sigma^2}\right)\binom{n}{k} q^k(1-q)^{n-k}.
\end{aligned} \label{discretegauss}
\end{equation}
As before, we can use a Gibbs sampler to simulate from the full conditionals of this kernel. For $q$, we draw an update from the Beta distribution as in 
\eqref{eq:fullcondq}. For $\hat{q}$, we evaluate \eqref{DGp} for all values $\hat{q} \in \{0, 1/n, 2/n, \dots, 1\}$ and normalize to obtain a probability mass function, from which we then sample to obtain the update. We construct the 95\%  interval using the $2.5$ percentile and  $97.5$ percentiles of the posterior draws of $q$. Alternatively, we can compute \eqref{discretegauss} for all $q \in Q_0$, and normalize the resulting kernels to approximate the posterior distribution over a grid of values for $p$.

\section{Simulations}\label{sec:sims}

In this section, we conduct simulation studies to examine the repeated sampling performances of the Wald-DP and Wilson-DP intervals from Section \ref{chap:failure} and the principled intervals from Section \ref{alternative}. 
We consider two sample sizes, $n = 100$ and $n=1000$, and four true proportions, $p \in \{0.1, 0.2, 0.5, 0.8\}$. 
For a given $n$ and $p$, in each simulation run we generate a value of $\hat p$ by sampling from a binomial distribution with $n$ trials and probability $p$. We then add noise to the sampled $\hat p$ using either a Laplace mechanism with $\epsilon \in \{0.1, 0.3, 0.5, 5\}$ or a discrete Gaussian mechanism with $\sigma^2 \in \{10, 3.33, 2, 0.2\}$.  When $\lambda=2$, these correspond to values of $\epsilon \in \{0.1, 0.3, 0.5, 5\}$, per the result described at the end of Section \ref{sec:renyi}.
For the intervals based on pure DP and the Laplace mechanism, we compute the intervals in Section \ref{alt_Bayes_flat} through Section \ref{alt_exact}.  For the intervals based on the discrete Gaussian mechanism and R\'enyi DP, we compute the interval in Section \ref{alt_renyi}. We illustrate the Bayesian posterior intervals using Gibbs samplers---so as to ensure continuous support for the posterior distribution---using $5000$ posterior draws.
We consider 95\% intervals for all methods. Finally, in each scenario we generate 5000 independent simulation runs.  Codes for all methods are available at 
\url{https://github.com/jkstatai/Interval_Estimation_Binomial_Proportion_DP}.

\begin{figure}[t]
    \centering
    \includegraphics[width=0.9\linewidth]{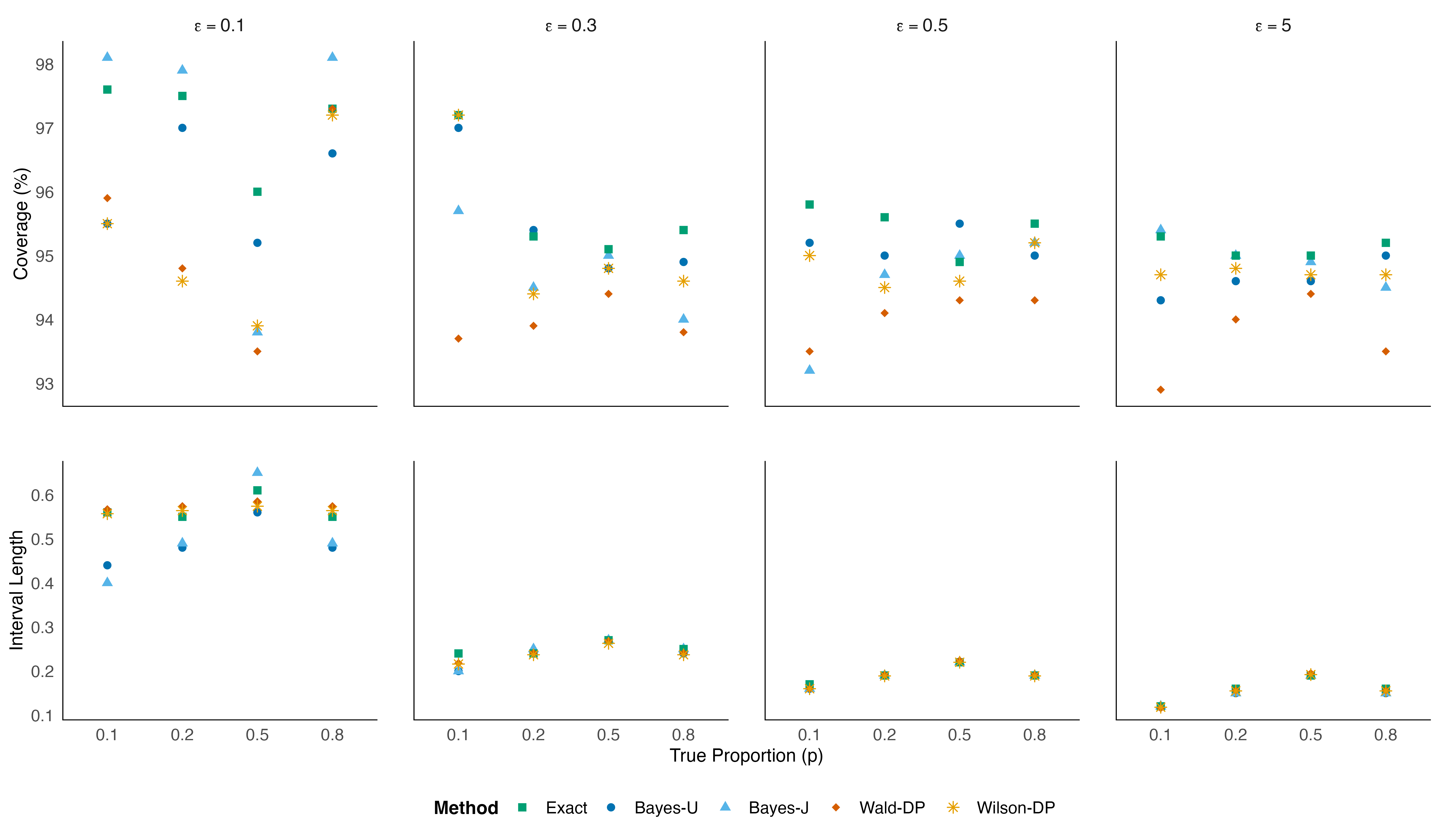}
    \caption[Coverage and average length of intervals under $\epsilon$-DP when $n=100$]{Coverage (\%) and average interval length for the Wald-DP, Wilson-DP, Bayesian credible intervals with uniform (Bayes-U) and Jeffreys (Bayes-J) priors, 
    and the exact interval (Exact) for $\epsilon$-DP with Laplace noise across different values of $\epsilon$ and $p$ when $n = 100$.}
    \label{fig:dp_coverage_length_n100}
\end{figure}

\begin{figure}[t]
    \centering
    \includegraphics[width=0.9\linewidth]{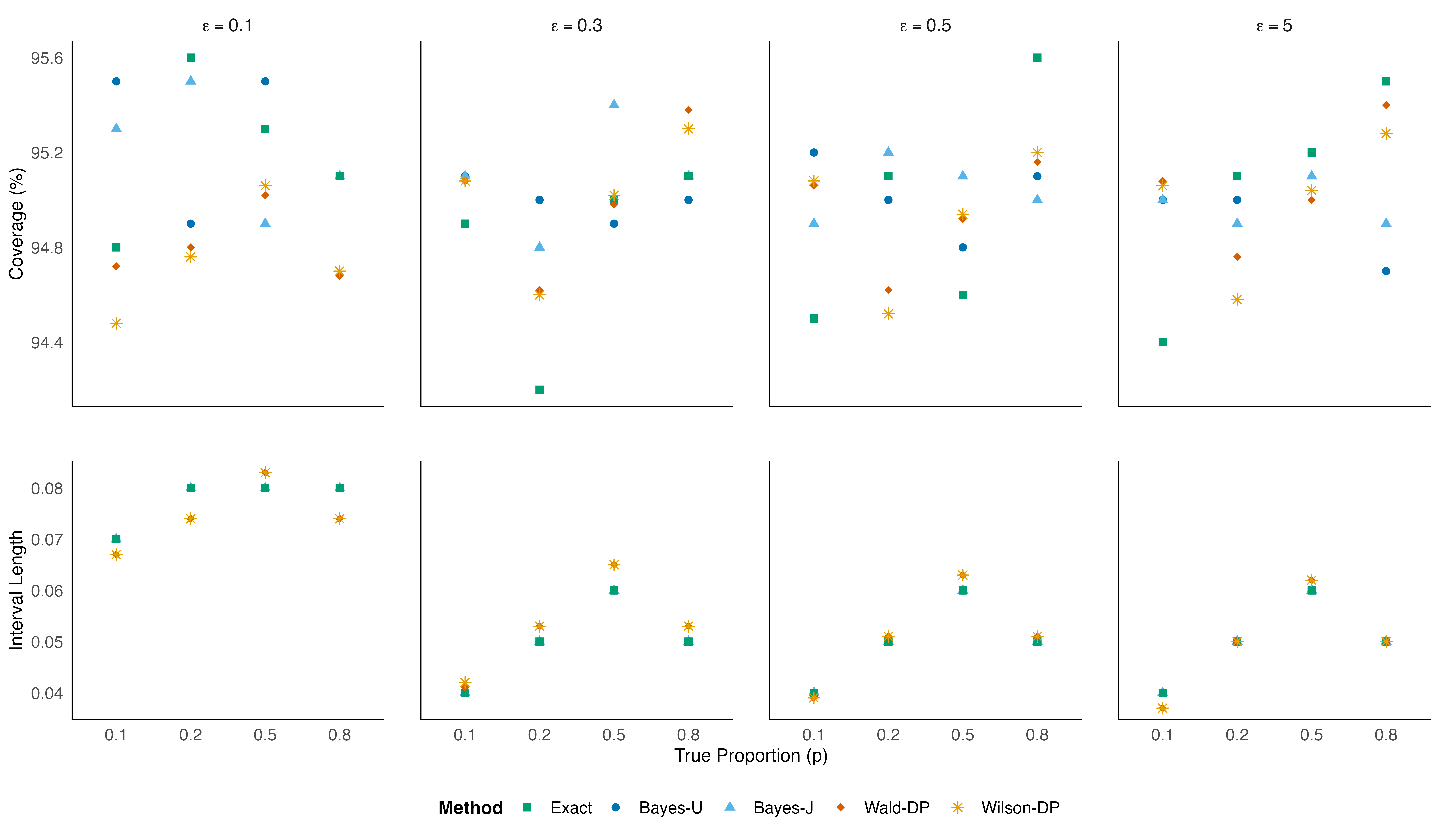}
    \caption[Coverage and average length of intervals under $\epsilon$-DP when $n=1000$]{Coverage (\%) and average interval length for the Wald-DP, Wilson-DP, Bayesian credible intervals with uniform (Bayes-U) and Jeffreys (Bayes-J) priors, 
    and the exact interval (Exact) for $\epsilon$-DP with Laplace noise across different values of $\epsilon$ and $p$ when $n = 1000$.}
    \label{fig:dp_coverage_length_n1000}
\end{figure}

Using this simulation design, we find that the Wald-DP interval frequently includes values outside $[0,1]$, especially for $p$ near the boundary when $n=100$ and $\epsilon<1$.  As examples, almost 90\% of the Wald-DP intervals are outside $[0,1]$ when $(p=0.1, n=100, \epsilon=0.1)$, and about 16\% of the intervals are outside $[0,1]$ when $(p=0.1, n=100, \epsilon=0.5)$. The Wilson-DP interval is  less prone to including values outside $[0,1]$, but it still happens quite often in these situations. Once $n=1000$, we find that the intervals rarely fall outside $[0,1]$ for both methods.  The supplementary material includes a table with the percentages of the 5000  Wald-DP and Wilson-DP intervals outside $[0,1]$ for each combination of $(p, n, \epsilon)$.  When comparing the repeated sampling properties of these intervals to the repeated sampling properties of the principled intervals, we clip the limits of the Wald-DP and Wilson-DP intervals to $[0,1]$.

Figure \ref{fig:dp_coverage_length_n100} and Figure \ref{fig:dp_coverage_length_n1000}  display the empirical coverage rates and average lengths when $n=100$ and $n=1000$, respectively, for the intervals based on the Wald-DP, Wilson-DP, and the principled methods under $\epsilon$-DP.  Overall, the coverage rates for all procedures are near the nominal 95\% rate.  However, we can discern some patterns.  
First, the empirical coverage rates for the Wald-DP (especially) and the Wilson-DP intervals often dip below the nominal 95\% rate, especially when $n=100$. They do so while also sometimes having larger average interval lengths than some of the principled methods.  Taken together, these results suggest the Wald-DP and Wilson-DP methods are not competitive methods.   Second, the exact and Bayesian intervals tend to perform similarly. They generally have coverage rates close to or exceeding the nominal 95\% rate, with exceptions in some scenarios. The most pronounced differences across these three methods appear when $(n=100, \epsilon = 0.1)$, as the exact intervals typically are wider than the Bayesian intervals.
This finding provides a rationale for preferring the Bayesian interval over the exact one, although using either is defensible.  Third, the Bayesian intervals with the uniform prior and a Jeffreys prior offer reasonably similar coverage rates and average lengths. This is expected, as the two intervals share nearly identical structures and use priors whose differences evidently do not strongly
influence the intervals in these simulations. The average lengths do differ visibly when $(n=100, \epsilon = 0.1)$, which is when we would expect the prior to have the most influence on the inferences. For example, in this setting, the Bayes-J interval tends to be narrower than the Bayes-U interval when $p=0.1$, whereas the opposite arises when $p=0.5$. Finally, also as expected, the average interval lengths decrease as $n$ and $\epsilon$ increase.  Indeed, when $(n=100, \epsilon = 0.1)$, the intervals tend to be so wide that they arguably do not locate $p$ with useful accuracy.  This is a price to pay for having such a strong privacy guarantee with a relatively small $n$.  For $n=1000$, the intervals tend to be narrow enough to locate $p$ even with small $\epsilon$.

\begin{figure}[t]
    \centering
    \includegraphics[width=0.9\linewidth]{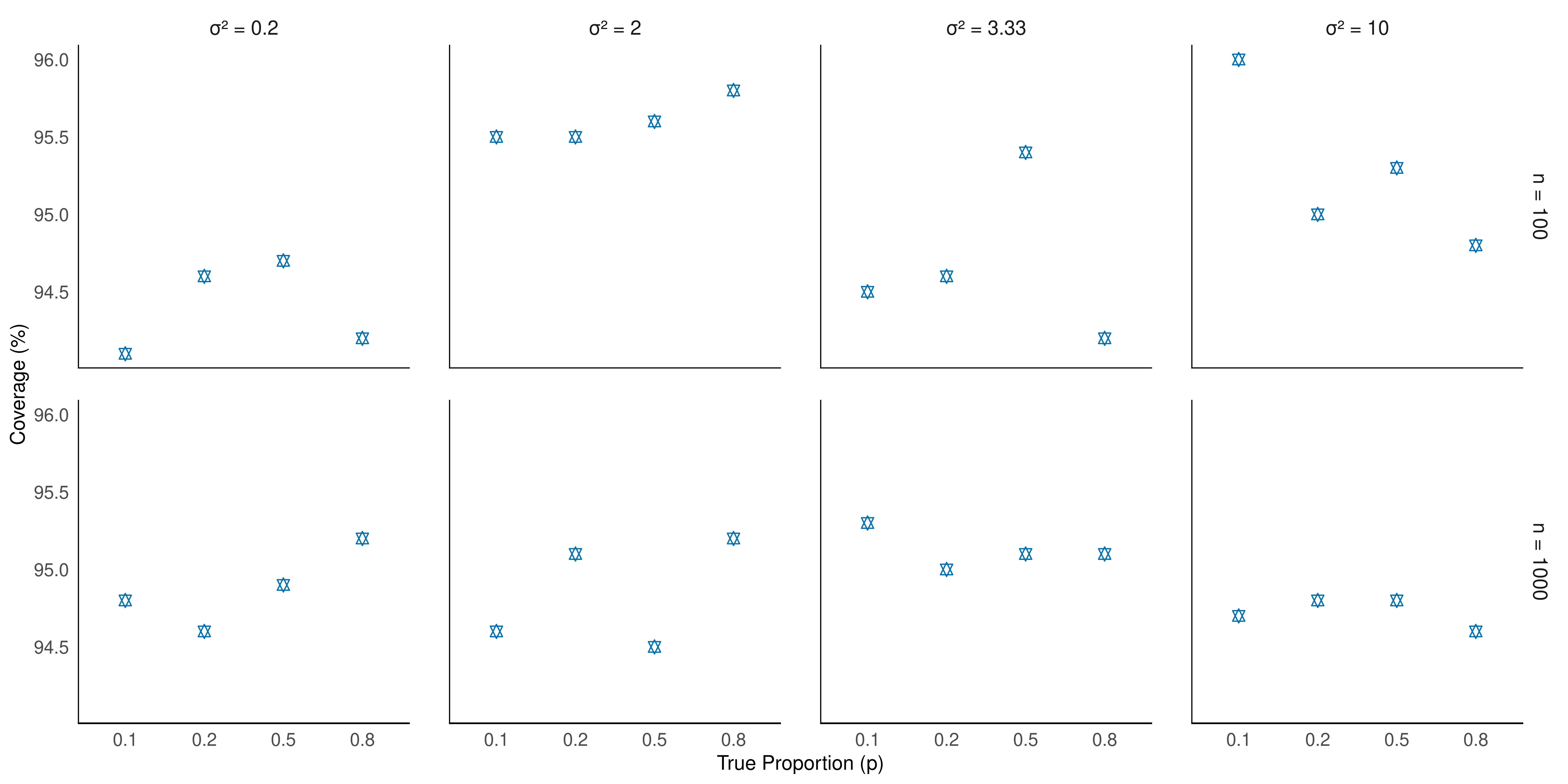}
    \caption[Coverage of Bayesian credible intervals under R\'enyi DP for $n=100$ and $n=1000$]{Coverage (\%) of Bayesian credible intervals under R\'enyi differential privacy using the discrete Gaussian mechanism for different values of $\sigma^2$ and true proportion $p$. Results are shown for $n = 100$ and $n = 1000$.}
    \label{fig:renyi_coverage_n100_1000}
\end{figure}

Turning to the R\'enyi DP intervals, Figure 
\ref{fig:renyi_coverage_n100_1000} displays the empirical coverage rates for the 5000 intervals for both $n=100$ and $n=1000$. 
Once again, the coverage rates tend to be near the nominal 95\% rate.  These results suggest that the post-processing strategy of computing Bayesian credible intervals can be effective for different types of DP mechanisms.

Finally, we note that tabular results for all methods in all simulations are provided in the supplementary material.

\section{Conclusions}\label{sec:conc}

To summarize, the simulations suggest that, among the interval estimates for a binomial $p$ considered here, the exact intervals and Bayesian credible intervals have the most desirable properties.  The Bayesian interval appears to have closer to nominal coverage rates when the DP noise substantially impacts the reported proportion.
The Wald-DP and Wilson-DP intervals can produce limits outside the feasible region of $[0,1]$, which makes them somewhat awkward to use.  Even with clipping the infeasible parts of the interval, they still can provide lower than nominal coverage rates.

In developing the intervals in Section \ref{alternative}, we presume that the agency releases the noisy proportion without post-processing.  Some agencies may prefer to release only a clipped version of the DP proportion, so as not to release any values outside $[0,1]$.  We can accommodate this setting by modifying the noise distribution to account for the truncation.  Specifically, the Bayesian intervals 
now use truncated Laplace distributions or truncated discrete Gaussian distributions for $f(\hat p^* | \hat q)$.  The exact intervals add a step when simulating $\hat q^*$, namely truncating any simulated values outside the limits to the closer of zero and one.  We conjecture that the truncation will result in wider interval estimates for all the principled methods, since the clipped version of $\hat p^*$ provides less information than $\hat p^*$ itself.

\section*{Conflict of Interest}
On behalf of all authors, the corresponding author states that there is no conflict of interest.


\bibliographystyle{agsm}
\bibliography{bibliography.bib}

@article{lottreiter2020wilson,
  author = {Anne Lott and Jerome P. Reiter},
  title = {Wilson confidence intervals for binomial proportions with multiple imputation for missing data},
  journal = {The American Statistician},
  volume = {74},
  number = {2},
  pages = {109--115},
  year = {2020},
}

@article{dwork:exposed,
author = {Dwork, C. and Smith, A. and  Steinke, T. and Ullman, J.},
title = {Exposed! A Survey of Attacks on Private Data},
journal = {Annual Review of Statistics and Its Application},
volume = 4,
year = 2017,
pages = {61--84}}

@article{dwork:roth,
author = {Dwork, Cynthia and Roth, Aaron},
title = {The Algorithmic Foundations of Differential Privacy},
year = {2014},
publisher = {Now Publishers Inc.},
volume = {9},
journal = {Foundations and Trends in Theoretical Computer Science},
pages = {211–-407}}

@techreport{census:recon,
author = {J. M. Abowd and T. Adams and R. Ashmead and D. Darais and S. Dey and S. L. Garfinkel and N. Goldschlag and D. Kifer and P. LeClerc and E. Lew and S. Moore and R. A. Rodriguez and R. N. Tadros and L. Vilhuber},
title = {A Simulated Reconstruction and Reidentification Attack on the 2010 {U.S.} Census: Full Technical Report},
year = 2023,
institution = {U. S. Bureau of the Census, working paper CES-23-63}}

@article{reiter2019differential,
  author = {Jerome P. Reiter},
  title = {Differential Privacy and Federal Data Releases},
  journal = {Annual Review of Statistics and Its Application},
  volume = {6},
  pages = {85--101},
  year = {2019},
  publisher = {Annual Reviews},
  address = {Palo Alto, CA},
}

@inproceedings{vu:slav,
author = {Vu, D. and Slavkovi\'c, A.},
year = {2009},
title  = {Differential privacy for clinical trial data: Preliminary evaluations},
booktitle = {IEEE International Conference on Data Mining Workshops},
pages = {138-–143},
publisher = {IEEE}}

@article{Wang2015RevisitingDP,
  title={Revisiting Differentially Private Hypothesis Tests for Categorical Data},
  author={Yue Wang and Jaewoo Lee and Daniel Kifer},
  journal={arXiv: Cryptography and Security},
  year={2015}
}

@inproceedings{DPchisq,
author = {Gaboardi, Marco and Lim, Hyun Woo and Rogers, Ryan and Vadhan, Salil}, 
title = {Differentially Private Chi-Squared Hypothesis Testing: Goodness of Fit and Independence Testing},
editors = {M. Balcan and K. Weinberger}, 
booktitle = {Proceedings of the 33rd International Conference on Machine Learning (ICML ‘16)},
publisher = {PMLR},
pages = {2111--2120},
year = 2016}

@article{awan2020differentially,
  author = {Jordan Awan and Aleksandra Slavkovi\'c},
  title = {Differentially Private Inference for Binomial Data},
  journal = {Journal of Privacy and Confidentiality},
  volume = {10},
  number = {1},
  pages = {725},
  year = {2020},
  publisher = {Carnegie Mellon University}
}

@inproceedings{dwork2006calibrating,
  author = {Cynthia Dwork and Frank McSherry and Kobbi Nissim and Adam Smith},
  title = {Calibrating Noise to Sensitivity in Private Data Analysis},
  booktitle = {Theory of Cryptography: Third Theory of Cryptography Conference, TCC 2006, New York, NY, USA, March 4-7, 2006. Proceedings},
  pages = {265--284},
  year = {2006},
  publisher = {Springer Berlin Heidelberg},
  address = {Berlin, Heidelberg}
}

@inproceedings{dinur:nissim,
author = {Irit Dinur and Kobbi Nissim},
year = 2003,
title = {Revealing information while preserving privacy},
booktitle = {Proceedings of the Twenty-second ACM SIGMOD-SIGACT-SIGART Symposium on Principles of Database Systems (PODS '03)},
publisher = {New York: ACM},
pages = {202-–210}}

@inproceedings{dwork2006differential,
  author = {Cynthia Dwork},
  title = {Differential Privacy},
  booktitle = {Automata, {L}anguages and {P}rogramming: 33rd {I}nternational {C}olloquium, {ICALP} 2006, {V}enice, {I}taly, July 10–14, 2006. {P}roceedings, {P}art II},
  editor = {Michele Bugliesi and Bart Preneel and Vladimiro Sassone and Ingo Wegener},
  pages = {1--12},
  year = {2006},
  publisher = {Springer},
  address = {Berlin}
}

@article{10.1214/ss/1009213286,
author = {Lawrence D. Brown and T. Tony Cai and Anirban DasGupta},
title = {{Interval estimation for a binomial proportion}},
volume = {16},
journal = {Statistical Science},
number = {2},
publisher = {Institute of Mathematical Statistics},
pages = {101 -- 133},
year = {2001}
}

@article{newcombe1998two,
  author = {Newcombe, Robert G.},
  title = {Two-sided confidence intervals for the single proportion: Comparison of seven methods},
  journal = {Statistics in Medicine},
  volume = {17},
  number = {8},
  pages = {857--872},
  year = {1998},
  month = {04},
  pmid = {9595616}
}

@misc{apple2017learning,
  author = {{Differential Privacy Team, Apple}},
  title = {Learning with Privacy at Scale},
  year = {2017},
  howpublished = {\url{https://docs-assets.developer.apple.com/ml-research/papers/learning-with-privacy-at-scale.pdf}}
}

@inproceedings {208167,
author = {Joe Near},
title = {Differential Privacy at Scale: Uber and {B}erkeley Collaboration},
booktitle = {Enigma 2018 (Enigma 2018)},
year = {2018},
address = {Santa Clara, CA},
publisher = {USENIX Association},
month = {01}
}

@inproceedings{DBLP:journals/corr/Mironov17,
author = {Mironov, I.},
title = {R\'enyi Differential Privacy},
booktitle = {2017 IEEE 30th Computer Security Foundations Symposium (CSF)}, 
year = 2017,
pages = {263--275}}

@article{Wallis01082013,
author = {Sean Wallis},
title = {Binomial Confidence Intervals and Contingency Tests: Mathematical Fundamentals and the Evaluation of Alternative Methods},
journal = {Journal of Quantitative Linguistics},
volume = {20},
number = {3},
pages = {178--208},
year = {2013},
publisher = {Routledge},
eprint = {https://doi.org/10.1080/09296174.2013.799918}
}

@article{karwa,
author = {Karwa, V. and Vadhan, S.},
year = 2017,
title = {Finite sample differentially private confidence intervals},
journal = {arXiv preprint},
pages = {arXiv:1711.03908}}

@techreport{honaker,
author = {D’{O}razio, V. and Honaker, J. and King, G.},
year = 2015,
title = {Differential privacy for social science inference},
institution = {Sloan Foundation Economics Research Paper, (2676160)}}

@article{li:reiter,
author = {Li, L. and Reiter, J. P.},
year = 2022, 
title = {Bayesian inference for estimating subset proportions using
differentially private counts}, 
journal = {Journal of Survey Statistics and Methodology}, 
volume = 10, 
pages = {785--803}}

@article{lin:bun,
author = {Shurong Lin and Mark Bun and Marco Gaboardi and Eric D. Kolaczyk and Adam Smith},
title = {Differentially private confidence intervals for proportions under stratified random sampling}, 
journal = {Electronic Journal of Statistics},
volume = 18, 
pages = {1455--1494}, 
year = 2024}

@inproceedings{kazan:reiter,
author = {Kazan, Z. and Reiter, J. P.},
year = 2024,
title= {Prior-itizing privacy: A {B}ayesian approach to setting the privacy budget in differential privacy},
editor = {A. Globerson and L. Mackey and D. Belgrave and A. Fan and U.
Paquet and J. Tomczak and C. Zhang},
booktitle = {Advances in Neural Information Processing Systems},
volume = 37, 
pages = {90384--90430}}

@incollection{zanellabéguelin2022bayesianestimationdifferentialprivacy,
  title = 	 {{B}ayesian Estimation of Differential Privacy},
  author =       {Zanella-B\'eguelin, Santiago and Wutschitz, Lukas and Tople, Shruti and Salem, Ahmed and R\"{u}hle, Victor and Paverd, Andrew and Naseri, Mohammad and K\"{o}pf, Boris and Jones, Daniel},
  booktitle = 	 {Proceedings of the 40th International Conference on Machine Learning},
  pages = 	 {40624--40636},
  year = 	 {2023},
  editor = 	 {Krause, Andreas and Brunskill, Emma and Cho, Kyunghyun and Engelhardt, Barbara and Sabato, Sivan and Scarlett, Jonathan},
  volume = 	 {202},
  series = 	 {Proceedings of Machine Learning Research},
  publisher =    {PMLR}}

@book{brent1973algorithms,
  author    = {Richard P. Brent},
  title     = {Algorithms for Minimization without Derivatives},
  publisher = {Prentice-Hall},
  address   = {Englewood Cliffs, NJ},
  year      = {1973},
  isbn      = {0-13-022335-2}
}

@article{Covington_2024,
   title={Unbiased Statistical Estimation and Valid Confidence Intervals Under Differential Privacy},
   ISSN={1017-0405},
   url={http://dx.doi.org/10.5705/ss.202022.0276},
   DOI={10.5705/ss.202022.0276},
   journal={Statistica Sinica},
   publisher={Statistica Sinica (Institute of Statistical Science)},
   author={Covington, Christian and He, Xi and Honaker, James and Kamath, Gautam},
   year={2024} }

@article{Canonne_2022,
   title={Discrete {G}aussian for Differential Privacy},
   volume={12},
   ISSN={2575-8527},
   url={http://dx.doi.org/10.29012/jpc.784},
   DOI={10.29012/jpc.784},
   number={1},
   journal={Journal of Privacy and Confidentiality},
   publisher={Journal of Privacy and Confidentiality},
   author={Canonne, Clement and Kamath, Gautam and Steinke, Thomas},
   year={2022},
   month=jul }

\clearpage
\renewcommand{\thetable}{S\arabic{table}}
\renewcommand{\thefigure}{S\arabic{figure}}
\setcounter{section}{0}
\renewcommand{\thesection}{S\arabic{section}}
\renewcommand{\thesubsection}{S\arabic{section}.\arabic{subsection}}

\clearpage
\begin{center}
    \textbf{\large Supplementary Material for Interval Estimation for Binomial Proportions Under Differential Privacy}
    
    \vspace{0.3cm}
    Hsuan-Chen (Justin) Kao \quad Jerome P. Reiter \\
    Department of Statistical Science, Duke University
\end{center}

\section{Introduction}
This document includes supplementary material for the main text.  In Section \ref{appendix:wilson_out_of_bound}, we show that the Wilson-DP interval is likely to produce limits that lie outside $[0,1]$.  In Section \ref{sec:tables}, we provide tabular summaries of the simulation results described in the main text.

\section{Bounds and the Wilson-DP Interval}
\label{appendix:wilson_out_of_bound}


In this section, we show that under pure DP with the Laplace mechanism, the Wilson-DP interval for $p$ easily can lead to interval bounds outside $[0, 1]$. To do so, we directly analyze the quadratic function derived from the Wilson interval under DP.

The inequality for the Wilson confidence interval under DP is
\begin{equation}
\left( \hat{p}^* - p \right)^2 \leq z_{\alpha/2}^2 \left( \frac{p(1 - p)}{n} + \frac{2}{n^2 \epsilon^2} \right),
\end{equation}
where $ \hat{p}^*$ represents the observed noisy proportion with Laplace noise added in and $p$  denotes the true proportion. The sample size is given by $n$, and $z_{\alpha/2}$ is the critical value from the standard normal distribution. Lastly, $\epsilon$ is the privacy parameter for the Laplace mechanism. 

Expanding both sides, we have  for the left-hand side 
    \begin{equation}\label{lhs}
    \left( \hat{p}^* - p \right)^2 = \hat{p}^{*2} - 2 \hat{p}^* p + p^2.
    \end{equation}
For the right-hand side, we have 
    \begin{equation}\label{rhs}
    z_{\alpha/2}^2 \left( \frac{p(1 - p)}{n} + \frac{2}{n^2 \epsilon^2} \right) = \frac{z_{\alpha/2}^2}{n} (p - p^2) + \frac{2 z_{\alpha/2}^2}{n^2 \epsilon^2}.
    \end{equation}
Subtracting \eqref{rhs} from \eqref{lhs}, we have 
\begin{equation} \label{wilson_inequality}
    \begin{aligned}
       p^2 + \frac{z_{\alpha/2}^2}{n} p^2 - \frac{z_{\alpha/2}^2}{n} p - 2 \hat{p}^* p + \hat{p}^{*2}  -\frac{2 z_{\alpha/2}^2}{n^2 \epsilon^2} & \leq 0 \\
       \left(1+\frac{z_{\alpha/2}^2}{n}\right)p^2 - \left(2 \hat{p}^* + \frac{z_{\alpha/2}^2}{n} \right)p + \left(\hat{p}^{*2}  -\frac{2 z_{\alpha/2}^2}{n^2 \epsilon^2} \right) & \leq 0 .
    \end{aligned}
\end{equation}
We multiply all terms by $n$ to match the definitions in the main text, so that 
\begin{equation}
    \left(n+z_{\alpha / 2}^2\right) p^2-
    \left(2 n \hat{p}^*+z_{\alpha / 2}^2\right) p+
    \left(n \hat{p}^{* 2}-\frac{2 z_{\alpha / 2}^2}{n \epsilon^2}\right) \leq 0.\label{eq:solvewilson}
\end{equation}
To solve \eqref{eq:solvewilson}, we separate the inequality into three parts, namely the  quadratic terms in \( p^2 \), 
the linear terms in \( p \),
and the constant terms $n \hat{p}^{* 2}-\frac{2 z_{\alpha / 2}^2}{n \epsilon^2}$.  We express the inequality  in \eqref{eq:solvewilson} in the standard quadratic form,
\begin{equation}
A p^2 + B p + C \leq 0,
\end{equation}
where 
$A = n + z_{\alpha/2}^2$,   
$B = - \left( 2 n \hat{p}^* + z_{\alpha/2}^2 \right)$, and
$C = n \hat{p}^{*2} - \frac{2 z_{\alpha/2}^2}{n \epsilon^2}$.
We use the quadratic formula to analyze its properties,
\begin{equation}
p = \frac{-B \pm \sqrt{B^2 - 4 A C}}{2 A},
\end{equation}
where the discriminant is
$D = B^2 - 4 A C.$
We consider the signs of the coefficients.  For $A$, we have 
    \begin{equation}
    A = n + z_{\alpha/2}^2 > 0. 
    \end{equation}
For $C$, we have
    \begin{equation}
    C = n \hat{p}^{*2} - \frac{2 z_{\alpha/2}^2}{n \epsilon^2}.
    \end{equation}
When \(\epsilon\) is small, \(2 z_{\alpha/2}^2/(n \epsilon^2) \) becomes large, potentially making \( C \) negative. Computing \( D \) to analyze the roots, we have 
\begin{align}
D &= B^2 - 4 A C \\
&= \left( - \left( 2 n \hat{p}^* + z_{\alpha/2}^2 \right) \right)^2 - 4 (n + z_{\alpha/2}^2) \left( n \hat{p}^{*2} - \frac{2 z_{\alpha/2}^2}{n \epsilon^2} \right) \\
&= \left( 2 n \hat{p}^* + z_{\alpha/2}^2 \right)^2 - 4 (n + z_{\alpha/2}^2) n \hat{p}^{*2} + \frac{8 z_{\alpha/2}^2 (n + z_{\alpha/2}^2)}{n \epsilon^2}\\
& =4 n \hat{p}^* z_{\alpha / 2}^2+z_{\alpha / 2}^4-4 n z_{\alpha / 2}^2 \hat{p}^{* 2}+\frac{8 z_{\alpha / 2}^2\left(n+z_{\alpha / 2}^2\right)}{n \epsilon^2} \\
& = 4 n z_{\alpha / 2}^2 \hat{p}^*\left(1-\hat{p}^*\right)+z_{\alpha / 2}^4+\frac{8 z_{\alpha / 2}^2\left(n+z_{\alpha / 2}^2\right)}{n \epsilon^2}.
\end{align}
The discriminant $D > 0$ ensures real roots exist. 

When we substitute these results into the expression, the roots of the quadratic equation are
\begin{equation}
p = \frac{2 n \hat{p}^* + z_{\alpha/2}^2 \pm \sqrt{D}}{2 (n + z_{\alpha/2}^2)} .
\end{equation}

We now demonstrate the out-of-bound issues for two extremes, $\hat{p}^* \rightarrow 0$ and $\hat{p}^* \rightarrow 1$.
\begin{itemize}
    \item Case 1: \( \hat{p}^* \rightarrow 0.\)
    When \( \hat{p}^* \) is close to 0:
    \begin{align}
    C &\approx - \frac{2 z_{\alpha/2}^2}{n \epsilon^2} < 0, \\
    B &\approx - z_{\alpha/2}^2 < 0, \\
    D &\approx z_{\alpha/2}^4 + \frac{8 z_{\alpha/2}^2 (n + z_{\alpha/2}^2)}{n \epsilon^2} > 0.
    \end{align}
    The lower root is 
    \begin{align}
    p_{\text{lower}} &= \frac{z_{\alpha/2}^2 - \sqrt{D}}{2 (n + z_{\alpha/2}^2)} < 0,
    \end{align}
since $\sqrt{D} > z_{\alpha/2}^2$.
    \item Case 2: \( \hat{p}^* \rightarrow 1.\)
    When \( \hat{p}^* \) is close to 1:
    \begin{align}
    C &\approx n - \frac{2 z_{\alpha/2}^2}{n \epsilon^2}, \\
    B &\approx - (2 n + z_{\alpha/2}^2 ) < 0, \\
    D &\approx z_{\alpha/2}^4 + \frac{8 z_{\alpha/2}^2 (n + z_{\alpha/2}^2)}{n \epsilon^2} > 0.
    \end{align}
    The upper root is
    \begin{align}
    p_{\text{upper}} &= \frac{2 n + z_{\alpha/2}^2 + \sqrt{D}}{2 (n + z_{\alpha/2}^2)} > 1,
    \end{align}
since the numerator exceeds the denominator.
    \end{itemize}

To sum up, the analysis shows that when $C < 0$, which is likely under small $\epsilon$, the roots of the quadratic equation can be outside $[0, 1]$. Specifically, when $ \hat{p}^*$ is close to 0, the lower limit of the Wilson-DP interval is likely negative; when $ \hat{p}^* $ is close to 1, the upper limit likely exceeds 1. 

\newpage
\section{Tabular Results from the Simulation Studies} \label{sec:tables}

Table \ref{tab:plugin_combined_n100_n1000} presents the results for the Wald-DP and Wilson-DP intervals.  Table \ref{tab:dp_all_methods_side_by_side} displays results used to make Figure 1 and Figure 2 in the main text. Table \ref{tab:renyi_dp_discrete_gaussian} includes the results used to make Figure 3 in the main text.

\begin{table}[h]
\caption[Coverage, average length (for $n=100$), and out-of-bound rates (for $n=100$ and $n=1000$) under $\epsilon$-DP]{
Coverage (\%) and average length for Wald-DP and Wilson-DP confidence intervals under $\epsilon$-DP with Laplace noise for $n=100$, and comparison of out-of-bound rates for both $n=100$ and $n=1000$, evaluated across different values of $\epsilon$ and $p$.}
\label{tab:plugin_combined_n100_n1000}
\centering
\scriptsize
\begin{tabular}{cc|cc|cc|cc|cc}
\hline
\multicolumn{2}{c|}{\textbf{Settings}} &
\multicolumn{2}{c|}{\textbf{Coverage (\%) ($n{=}100$)}} &
\multicolumn{2}{c|}{\textbf{Average Length ($n{=}100$)}} &
\multicolumn{2}{c|}{\textbf{Out-of-Bound ($n{=}100$)}} &
\multicolumn{2}{c}{\textbf{Out-of-Bound ($n{=}1000$)}} \\
$\epsilon$ & $p$ & Wald-DP & Wilson-DP & Wald-DP & Wilson-DP & Wald-DP & Wilson-DP & Wald-DP & Wilson-DP \\
\hline
0.1 & 0.1 & 95.9 & 95.5 & .567 & .557 & .898 & .884 & .001 & .0006 \\
0.1 & 0.2 & 94.8 & 94.6 & .574 & .564 & .750 & .727 & .000 & .0000 \\
0.1 & 0.5 & 93.5 & 93.9 & .585 & .574 & .143 & .125 & .000 & .0000 \\
0.1 & 0.8 & 97.3 & 97.2 & .574 & .564 & .782 & .747 & .000 & .0000 \\
0.3 & 0.1 & 93.7 & 97.2 & .218 & .216 & .597 & .445 & .000 & .0000 \\
0.3 & 0.2 & 93.9 & 94.4 & .241 & .237 & .067 & .040 & .000 & .0000 \\
0.3 & 0.5 & 94.4 & 94.8 & .268 & .263 & .000 & .000 & .000 & .0000 \\
0.3 & 0.8 & 93.8 & 94.6 & .241 & .237 & .069 & .040 & .000 & .0000 \\
0.5 & 0.1 & 93.5 & 95.0 & .160 & .160 & .273 & .128 & .000 & .0000 \\
0.5 & 0.2 & 94.1 & 94.5 & .191 & .189 & .008 & .002 & .000 & .0000 \\
0.5 & 0.5 & 94.3 & 94.6 & .224 & .220 & .000 & .000 & .000 & .0000 \\
0.5 & 0.8 & 94.3 & 95.2 & .191 & .189 & .007 & .002 & .000 & .0000 \\
5.0 & 0.1 & 92.9 & 94.7 & .116 & .118 & .010 & .000 & .000 & .0000 \\
5.0 & 0.2 & 94.0 & 94.8 & .156 & .155 & .000 & .000 & .000 & .0000 \\
5.0 & 0.5 & 94.4 & 94.7 & .195 & .192 & .000 & .000 & .000 & .0000 \\
5.0 & 0.8 & 93.5 & 94.7 & .156 & .155 & .000 & .000 & .000 & .0000 \\
\hline
\end{tabular}
\end{table}

\begin{table}[H]
\caption[Results for Bayesian credible interval performance for $n=100$ and $n=1000$ under $\epsilon$-DP]{Coverage (\%) and average length of Bayesian credible intervals with uniform (Bayes-U) and Jeffreys (Bayes-J) priors, and the exact interval (Exact) for $\epsilon$-DP with Laplace noise for $n=100$ and $n=1000$.}
\label{tab:dp_all_methods_side_by_side}
\centering
\scriptsize
\begin{tabular}{cc|cccccc|cccccc}
\hline
\multicolumn{2}{c|}{\textbf{Settings}} &
\multicolumn{6}{c|}{\textbf{Coverage (\%)}} &
\multicolumn{6}{c}{\textbf{Average Length}} \\
$\epsilon$ & $p$ &
\multicolumn{2}{c}{Bayes-U} &
\multicolumn{2}{c}{Bayes-J} &
\multicolumn{2}{c|}{Exact} &
\multicolumn{2}{c}{Bayes-U} &
\multicolumn{2}{c}{Bayes-J} &
\multicolumn{2}{c}{Exact} \\
& &
100 & 1000 & 100 & 1000 & 100 & 1000 &
100 & 1000 & 100 & 1000 & 100 & 1000 \\
\hline
0.1 & 0.1 & 95.5 & 95.5 & 98.1 & 95.3 & 97.6 & 94.8 & .44 & .07 & .40 & .07 & .56 & .07 \\
0.1 & 0.2 & 97.0 & 94.9 & 97.9 & 95.5 & 97.5 & 95.6 & .48 & .08 & .49 & .08 & .55 & .08 \\
0.1 & 0.5 & 95.2 & 95.5 & 93.8 & 94.9 & 96.0 & 95.3 & .56 & .08 & .65 & .08 & .61 & .08 \\
0.1 & 0.8 & 96.6 & 95.1 & 98.1 & 95.1 & 97.3 & 95.1 & .48 & .08 & .49 & .08 & .55 & .08 \\
0.3 & 0.1 & 97.0 & 95.1 & 95.7 & 95.1 & 97.2 & 94.9 & .20 & .04 & .20 & .04 & .24 & .04 \\
0.3 & 0.2 & 95.4 & 95.0 & 94.5 & 94.8 & 95.3 & 94.2 & .24 & .05 & .25 & .05 & .24 & .05 \\
0.3 & 0.5 & 94.8 & 94.9 & 95.0 & 95.4 & 95.1 & 95.0 & .27 & .06 & .27 & .06 & .27 & .06 \\
0.3 & 0.8 & 94.9 & 95.0 & 94.0 & 95.1 & 95.4 & 95.1 & .24 & .05 & .25 & .05 & .25 & .05 \\
0.5 & 0.1 & 95.2 & 95.2 & 93.2 & 94.9 & 95.8 & 94.5 & .16 & .04 & .16 & .04 & .17 & .04 \\
0.5 & 0.2 & 95.0 & 95.0 & 94.7 & 95.2 & 95.6 & 95.1 & .19 & .05 & .19 & .05 & .19 & .05 \\
0.5 & 0.5 & 95.5 & 94.8 & 95.0 & 95.1 & 94.9 & 94.6 & .22 & .06 & .22 & .06 & .22 & .06 \\
0.5 & 0.8 & 95.0 & 95.1 & 95.2 & 95.0 & 95.5 & 95.6 & .19 & .05 & .19 & .05 & .19 & .05 \\
5.0 & 0.1 & 94.3 & 95.0 & 95.4 & 95.0 & 95.3 & 94.4 & .12 & .04 & .12 & .04 & .12 & .04 \\
5.0 & 0.2 & 94.6 & 95.0 & 95.0 & 94.9 & 95.0 & 95.1 & .15 & .05 & .15 & .05 & .16 & .05 \\
5.0 & 0.5 & 94.6 & 95.2 & 94.9 & 95.1 & 95.0 & 95.2 & .19 & .06 & .19 & .06 & .19 & .06 \\
5.0 & 0.8 & 95.0 & 94.7 & 94.5 & 94.9 & 95.2 & 95.5 & .15 & .05 & .15 & .05 & .16 & .05 \\
\hline
\end{tabular}
\end{table}

\begin{table}[H]
\caption[Coverage and average length of Bayesian credible intervals with discrete Gaussian noise under R\'enyi DP ($\alpha=2, 10$)]{Coverage (\%) and average interval length of Bayesian credible intervals under R\'enyi differential privacy
using the discrete Gaussian mechanism, evaluated across different values of $\sigma^2$ 
and $p$ for sample sizes $n=100$ and $n=1000$.}
\label{tab:renyi_dp_discrete_gaussian}
\centering
\scriptsize
\begin{tabular}{cc|cc|cc}
\hline
\multicolumn{2}{c|}{\textbf{Settings}} &
\multicolumn{2}{c|}{\textbf{$n=100$}} &
\multicolumn{2}{c}{\textbf{$n=1000$}} \\
$\sigma^2$ & $p$ &
\shortstack{Coverage (\%)} &
\shortstack{Average Length} &
\shortstack{Coverage (\%)} &
\shortstack{Average Length}
\\
\hline
10 & 0.1 & 96.0 & .16 &  94.7 & .04 \\
10 & 0.2 & 95.0 & .20 &  94.8 & .05 \\
10 & 0.5 & 95.3 & .23 &  94.8 & .06 \\
10 & 0.8 & 94.8 & .20 &  94.6 & .05 \\
$3.33$ & 0.1 & 94.5 & .14 &  95.3 & .04 \\
$3.33$ & 0.2 & 94.6 & .17 &  95.0 & .05 \\
$3.33$ & 0.5 & 95.4 & .20 &  95.1 & .06 \\
$3.33$ & 0.8 & 94.2 & .17 &  95.1 & .05 \\
$2$ & 0.1 & 95.5 & .13 &  94.6 & .04 \\
$2$ & 0.2 & 95.5 & .16 &  95.1 & .05 \\
$2$ & 0.5 & 95.6 & .20 &  94.5 & .06 \\
$2$ & 0.8 & 95.8 & .16 &  95.2 & .05 \\
$0.2$ & 0.1 & 94.1 & .12 &  94.8 & .04 \\
$0.2$ & 0.2 & 94.6 & .15 &  94.6 & .05 \\
$0.2$ & 0.5 & 94.7 & .19 &  94.9 & .06 \\
$0.2$ & 0.8 & 94.2 & .15 &  95.2 & .05 \\
\hline
\end{tabular}
\end{table}

\end{document}